\documentclass[%
 reprint,
 superscriptaddress,
 amsmath,amssymb,
 aps,
 prx
]{revtex4-2}

\usepackage{braket}
\usepackage{graphicx}
\usepackage{dcolumn}
\usepackage{bm}
\usepackage{siunitx}
\usepackage{xcolor} 
\usepackage{tabularx}
\usepackage{dsfont}					
\usepackage{mathrsfs}				
\usepackage{float}
\usepackage{hyperref}

\newcommand{\be}[0]{\begin{equation}}	
\newcommand{\ee}[0]{\end{equation}}

\usepackage{hyperref}
\hypersetup{colorlinks, 
	linkcolor={blue!75!black!80!yellow},
	citecolor={blue!75!black!80!yellow}, 
	urlcolor={blue!75!black!80!yellow}
	}

\usepackage[normalem]{ulem}

\begin{document}


\title{Absence of Weak Localization on Negative Curvature Surfaces}


\author{Jonathan B. Curtis}
\author{Prineha Narang}
\email[]{joncurtis@ucla.edu}
\affiliation{College of Letters and Science, University of California, Los Angeles, CA 90095, USA}
\author{Victor Galitski}
\affiliation{Joint Quantum Institute, University of Maryland, College Park, MD 20783 USA}

\date{\today}

\begin{abstract}
The interplay between  disorder and quantum interference leads to a wide variety of physical phenomena including celebrated Anderson localization -- the complete absence of diffusive transport due to quantum interference between different particle trajectories. 
In two dimensions, any amount of disorder is thought to induce  localization of all states at long enough length scales, though this may be prevented if bands are topological or have strong spin-orbit coupling. In this note, we present a simple argument providing another mechanism for disrupting localization: by tuning the underlying curvature of the manifold on which diffusion takes place.  We show that negative curvature manifolds contain a natural infrared cut off for the probability of self-returning paths. We provide explicit calculations of the Cooperon -- directly related to the weak-localization corrections to the conductivity -- in hyperbolic space. It is shown that constant negative curvature leads to a rapid growth in the number of available trajectories a particle can coherently traverse in a given time, reducing the importance of interference effects and restoring classical diffusive behavior even in the absence of inelastic collisions.  We conclude by arguing that this result may be amenable to experimental verification through the use of quantum simulators.
\end{abstract}

\maketitle
The interplay between quantum mechanics and geometry is rich and driven by deep questions ranging from the nature of spacetime to quantum chaos~\cite{Maldacena.1999,Ryu.2006,HAWKING.1974,Hawking.1975}.
Equally rich is the interplay between  disorder and quantum mechanics.  Whereas in classical systems, disorder invariably leads to the diffusive motion of particles, in quantum systems interference effects can  suppress diffusion altogether, rendering particles localized in real-space~\cite{Anderson.1956,ALTSHULER.1985}. 

In two dimensions, a simple scaling argument~\cite{Abrahams.1979} indicates that Anderson localization will onset the moment any disorder is introduced. However, in recent decades it has been shown that there are a few ways a system may escape this fate.
A non-trivial band topology~\cite{Fendley.2000} is one such way, leading to the emergence of stable integer quantum Hall plateaus~\cite{Pruisken.1988,Pruisken.1989}.  Strong spin-orbit coupling can also disrupt localization physics, leading to the weak anti-localization effect instead~\cite{Efetov.1980,Lee.1985oxt}. 

\begin{figure}
    \centering
    \includegraphics[width=\linewidth]{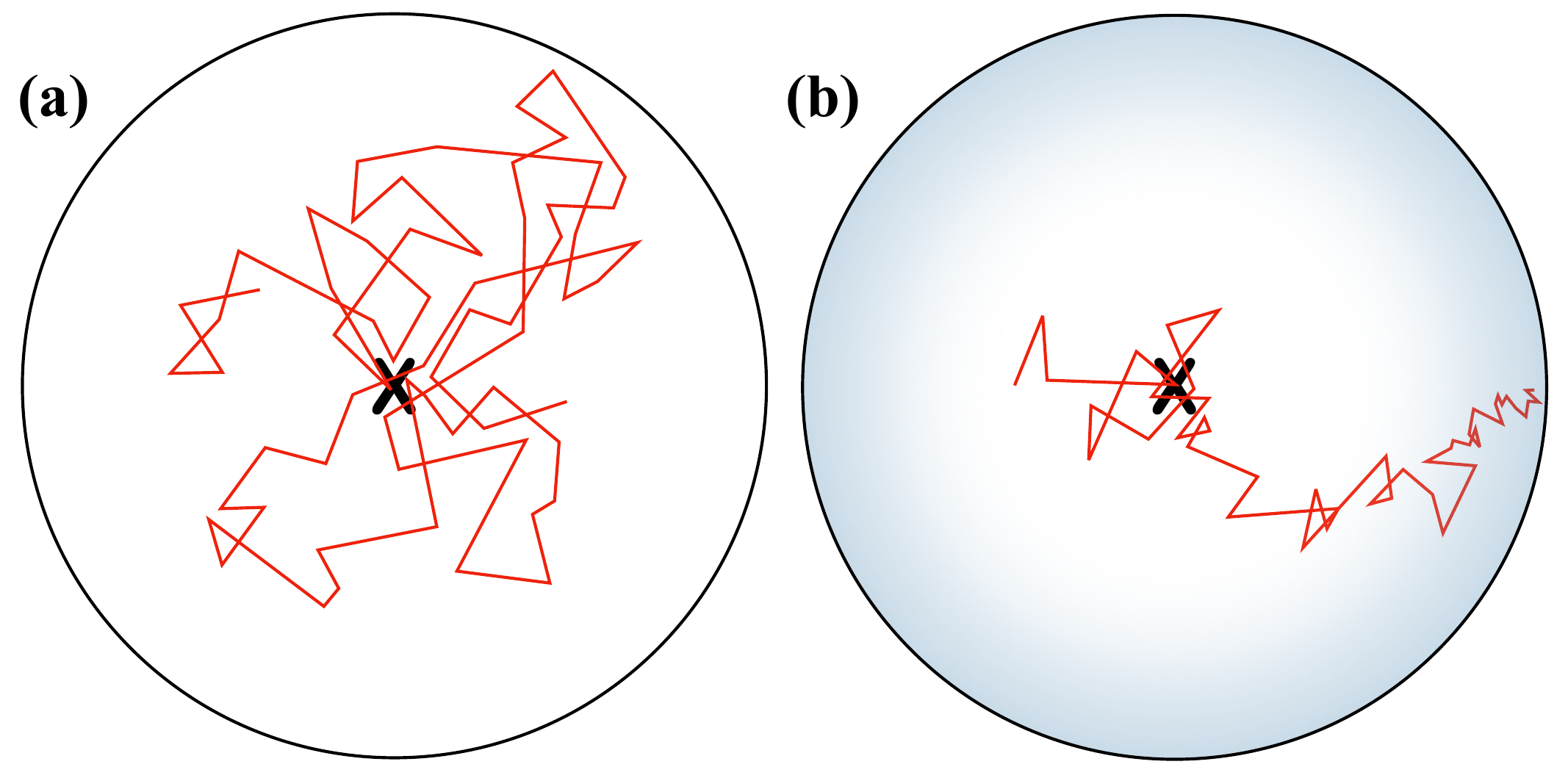}
    \caption{(a) Schematic illustration of random walk trajectories in {\bf flat} space. 
    As a particle undergoes diffusion away from the initial impurity scattering event (illustrated by the X), it randomly walks on the manifold. 
    In flat space these trajectories will always return to the origin after a long enough time, implying that self-intersecting trajectories will always be present, and their resulting interference will produce a divergent weak-localization correction. 
    (b) Schematic illustration of random-walk trajectories in {\bf hyperbolic} space, with curvature illustrated as the onset of the coloration away from the origin.
    As a particle undergoes diffusion away from the initial impurity scattering event (illustrated by the X), it randomly walks on the manifold.    
    Unlike in flat space, due to the onset of negative curvature, these trajectories will eventually wander away from the origin and never return.
    This implies that at long times self-intersecting trajectories are irrelevant, and the weak-localization corrections remain finite at low frequencies.
    }
    \label{fig:schematic}
\end{figure}

A particularly intuitive approach to understand localization physics in a weak disorder potential was proposed by Althsuler and Aronov~\cite{ALTSHULER.1985}, and is rooted in the iconic Feynman path integral theory. The latter transplants classical intuition to quantum mechanics by interpreting the quantum mechanical transition amplitudes through a sum over all classical trajectories ``weighted'' with $e^{iS/\hbar}$, where $S$ is the classical action for each path. Many canonical quantum mechanical problems can be interpreted in a different light through this Lagrangian approach. For example, the quasiclassical approximation corresponds to paths confined within a thin ``tube'' surrounding the unique classical trajectory, tunneling phenomena can be reinterpreted in terms of instantons, and the existence (1D and 2D) or absence (3D and above) of bound states in a weak potential can be tied one-to-one to the probability of self-return of a diffusive particle. The Althsuler-Aronov argument for weak-localization is similar to the latter in that it ties the infrared weak localization divergence to the probability of self-return for diffusion in dimensions  two and below. Our key argument for the absence of localization in hyperbolic space can be summarized in one sentence: in contrast to flat space, a diffusive particle on a two-dimensional surface with negative curvature is not guaranteed to return to the origin. 
This is summarized schematically in Fig.~\ref{fig:schematic}.

Below we elaborate on this observation on the example of the hyperbolic space. The system we have in mind is a system of non-interacting fermions moving in two dimensions on a manifold with constant negative curvature. We will not consider the many-body version of the problem and focus on single-particle diffusion instead. Note also that the interaction effects in the conventional theory of weak localization are significant in that they provide a temperature-dependent  infrared cut off (dephasing). Since, as we show below, the diffusion kernel is free from any infrared divergences in the hyperbolic space, the interaction effects are less relevant. 
%

Consider a manifold with metric tensor $g_{ab}$ such that the invariant line-element is 
\begin{equation}
ds^2 = g_{ab} dx^a dx^b = dr^2 +R^2 \sinh^2 (r/R) d\phi^2 .
\end{equation}
This describes the Poincar{\'e} disk model of hyperbolic geometry with negative curvature $\kappa = -1/R^2$ and radius of curvature $R$~\footnote{One may take the limit of $r \ll R$ to recover the flat space (parameterized in polar coordinates), or analytically continue $R \to iR$ to obtain a spherical manifold with positive curvature $\kappa = -1/(iR)^2 >0$.}.
Consider a particle which moves diffusively with diffusion constant $D$ such that within a time $t$ we expect the particle to have randomly traversed a linear distance $L \sim \sqrt{Dt}$.
By integrating the area form $dA = d^2 x \sqrt{g}$ up to proper radius $L$, we see that the area enclosed by this trajectory is 
\begin{equation}
    A = 2\pi R^2 \left( \cosh \frac{L}{R} - 1\right).
\end{equation}
Note that for small radii $L\ll R$ this recovers to the usual formula $A = \pi L^2$, but for larger radii this grows as $\sim R^2 e^{L/R}$. 
We can then invert this formula to find out how much time is required for the particle's trajectory to enclose an area $A$, yielding 
\begin{equation}
    t = \frac{R^2}{D} \textrm{arcosh}^2\left( 1 + \frac{A}{2\pi R^2} \right).
\end{equation}
If we set the area equal to the sample size this essentially yields the Thouless time required to explore the entire system. 
Whereas in flat space, this time grows linearly with the system size, in a hyperbolic geometry this only grows as $\log^2 A$ for large systems, implying that very rapidly the particle can diffuse across the entire sample and therefore become an extended state. 


To be a bit more quantitative, we turn to the weak-localization quantum corrections to the conductivity in the two-dimensional curved space. The classical Drude conductivity can be written as $\sigma_0(0) = \nu_F D$, where $\nu_F$ is the density of states at the Fermi level  and $D = \frac{v_F^2 \tau_{\rm imp}}{2}$ is the diffusion constant in terms of the elastic impurity scattering time $\tau_{\rm imp}$ and Fermi velocity $v_F$.
Beyond leading order there are quantum correction due to ``maximally-crossed diagrams," which correspond to coherent backscattering of particles off of impurities, and are evaluated in terms of particle-particle correlation function, or Cooperon $\mathcal{C}(\mathbf{r},\mathbf{r}'; t)$.
This describes the transition amplitude for a pair of particles to traverse time-reversal partner trajectories connecting $\mathbf{r}'$ to $\mathbf{r}$ in a time $t$.  

In flat space, the weak-localization correction to the conductivity~\cite{ALTSHULER.1985,Lee.1985oxt} is given in terms of the Cooperon as 
\begin{equation}
\label{eqn:wl-correction}
    \frac{\delta \sigma_{\rm WL}(\omega)}{\sigma_0(0)} = - \tau^2_{\rm imp} \int dt e^{i\omega t}\mathcal{C}(\mathbf{r},\mathbf{r};t).
\end{equation}
At low frequencies this is related to the return-probability for the particle undergoing diffusion in the self-intersecting trajectory. This can be seen by noting that in the absence of time-reversal symmetry breaking, the Cooperon obeys a diffusion equation which reads 
\begin{equation}
    \left[ \frac{\partial}{\partial t} - D \nabla^2_{\mathbf{r}} \right]\mathcal{C}(\mathbf{r},\mathbf{r}';t) = \frac{1}{2\pi \nu_F \tau_{\rm imp}^2}  \delta^2(\mathbf{r}-\mathbf{r}') \delta(t).
\end{equation}
This is valid within the quasiclassical approximation, which is justified by the hierarchy of length scales $k_F^{-1}\ll \ell_{\rm imp}$, where $\ell_{\rm imp} = \tau_{\rm imp} v_F$ is the impurity scattering mean-free path, and $k_F$ is the Fermi momentum. 

In Euclidean space, the diffusion kernel can be obtained easily in momentum and frequency space, yielding $\mathcal{C}(\mathbf{q},\omega) \sim \frac{1}{2\pi \nu_F \tau_{\rm imp}^2}\frac{1}{D\mathbf{q}^2 -i\omega}$. 
We can then perform the integral, implementing a short- and a long-time cutoffs to obtain $\delta \sigma_{\rm WL}(0)/\sigma_0(0) \sim  -\log \frac{\tau_{\phi}}{\tau_{\rm imp}}$.
The upper (ultraviolet) cutoff at $\omega \sim \tau_{\rm imp}^{-1}$ is not important, as it simply corresponds to short-time deviations of the Cooperon dynamics from diffusion which only sets in at long times.
However, the lower (infrared) cutoff at the dephasing time $\omega \sim \tau_{\phi}^{-1}$ is important.
If we probe the DC conductivity correction, then this logarithmically diverges as $\log \tau_\phi/\tau_{\rm imp}$ as the dephasing time grows with the lowering of the temperature~\cite{ALTSHULER.1985}.
As a result, at some low-frequency comparable to either the temperature or phase-breaking rate, the quantum corrections become arbitrarily large, signaling the onset of Anderson localization once $\delta \sigma_{\rm WL}(0)/\sigma_0(0)\sim -1$. 

In order to study the analogue of localization in a curved geometry, we must first promote the diffusion equation to living on a curved manifold. Given the metric tensor $g_{ab}$, the appropriate generalization of the Laplacian operator ${\bm\nabla}^2 \to \frac{1}{\sqrt{g}} \partial_a g^{ab}\sqrt{g} \partial_b$ where $g = \det(g_{ab})$ and  $g^{ab}$ is the inverse of the metric tensor. 
The corresponding Cooperon equation on the curved surface then reads 
\begin{equation}
\label{eqn:diffusion-curved}
    \left[ \frac{\partial}{\partial t} - D \frac{1}{\sqrt{g}} \partial_a g^{ab}\sqrt{g} \partial_b \right]\mathcal{C}(\mathbf{r},\mathbf{r}';t) =  \zeta \delta^2(\mathbf{r},\mathbf{r}') \delta(t),
\end{equation}
with suitable generalization of the Dirac delta function $\delta^2(\mathbf{r},\mathbf{r}')$ to curved space. 
In flat space, the constant $\zeta = \frac{1}{2\pi \nu_F \tau_{\rm imp}^2}$, though we recognize that the microscopic derivation of this prefactor may be different in curved space so we allow for a generalization to parameter $\zeta$ provided it matches the correct units.
We expect that our covariant approach will certainly breakdown once the radius of curvature is no longer the longest length scale, which means we expect this approach to be valid only if $R \gg \ell_{\rm imp} \gg k_F^{-1}$.

In general, Eq.~\eqref{eqn:diffusion-curved} may be very difficult to solve for a complicated geometry, but luckily in the case of uniform negative curvature, the solution is known~\cite{Grigor'yan.1998} and is given by 
\begin{equation}
   \mathcal{C}(\mathbf{x},t)  = \zeta \frac{R e^{- D|t|/(4 R^2)}}{\sqrt{32 \pi^3 (D|t|)^3 }} \int_{x/R}^{\infty} \frac{ds s e^{- R^2 s^2/(4 D|t|)}}{\sqrt{\cosh s - \cosh(x/R)}}
\end{equation}
where $x = |\mathbf{x}| = | \mathbf{r}-\mathbf{r}'|$ is the invariant distance away from the initial point of the propagator. 
As a sanity check, if we take the limit of $\sqrt{Dt}, x\ll R$, such that the radius of curvature becomes very large this recovers the flat space limit.

We now evaluate the weak-localization correction using the curved space analogue of Eq.~\eqref{eqn:wl-correction}
\begin{multline}
    \delta \sigma_{\rm WL}(0)/\sigma_0(0) = - \tau_{\rm imp}^2 \mathcal{C}(0,0) \\
    = - \tau_{\rm imp}^2\zeta   \int_{\tau_{\rm imp}}^{\tau_\phi}dt  \frac{R e^{- Dt/(4 R^2)}}{\sqrt{32 \pi^3 (Dt)^3 }} \int_{0}^{\infty} \frac{ds s e^{- R^2 s^2/(4 Dt)}}{\sqrt{\cosh s - 1}}.
\end{multline}
The integral can be roughly evaluated by breaking in to two regimes: $t \ll R^2/D \equiv \tau_R$ and $t \gg \tau_R$, assuming that we are in the interesting regime where $\tau_\phi \gg \tau_R$.
For $t \ll \tau_R$ we can approximate the integral over $s$ by $\sqrt{2} \int_0^\infty ds e^{-s^2\tau_R/(4t)}$ since in this case the Gaussian will be sharply peaked around $s = 0$ so that $\cosh s = 1 + s^2/2$.
The integral is then carried out over $t$, integrating up to $\tau_R$.
This then contributes a term of order $\delta \sigma_{\rm WL}(0)/\sigma_0(0) \sim -\tau_{\rm imp}^2 \zeta \int_{\tau_{\rm imp}}^{\tau_R} \frac{dt}{4 \pi Dt}$. 

The second contribution comes from long times with $t \gg \tau_R$. 
In this case, we may instead evaluate the integral over $s$ by dropping the broad Gaussian altogether, evaluating $\int_{0}^{\infty} \frac{ds s }{\sqrt{\cosh s - 1}} = \pi^2/\sqrt{2}$.
Then the integral over time can be evaluated simply as $\delta \sigma_{\rm WL}(0)/\sigma_0(0) \sim -\tau_{\rm imp}^2 \zeta \frac{R\sqrt{\pi}}{8 D^{\frac32}} \int_{\tau_R}^{\tau_\phi} dt e^{-t/4\tau_R}/t^{\frac32}$.
This integral converges exponentially fast, so we take $\tau_\phi$ to infinity and find it simply gives a numerical contribution of order $-.15 \tau_{\rm imp}^2 \zeta/D $, and therefore simply renormalizes the exact value of $\tau_R$ taken in the logarithmic cutoff.

We thus find the conductance correction 
\begin{equation}
    \delta \sigma_{\rm WL}(0)/\sigma_0(0) \sim - \frac{ \tau_{\rm imp}^2\zeta }{4\pi D}\log\left(\frac{\tau_R}{\tau_{\rm imp}}\right).
\end{equation}
As a result, we see that the presence of the negative spatial curvature has suppressed the weak-localization correction and in particular, the presence of underlying spatial curvature has effectively regularized the infrared divergence which presents in flat space, replacing $\tau_\phi$ by the new time-scale $\tau_R$.

To conclude, we have argued that the underlying geometry of a manifold can influence quantum localization effects on the surface, and in particular negative spatial curvature suppresses localization effects.
This was based on a heuristic argument about the scaling of the Thouless time, but a more sophisticated calculation showed indeed that the Cooperon corrections remain finite at low-frequencies due to the spatial curvature, and this indicates that localization may only onset at a finite disorder strength on a curved surface. 

We also point out that the same kind of integral we encounter here occurs in many other contexts, including the simple single-particle problem of a bound state in a weak potential well, and the Mermin-Wagner theorem which forbids the emergence of broken-symmetry ordered phases in two dimensions. By the same argument as we have presented here, these familiar results in flat geometry are expected to be modified in hyperbolic space.

One can also imagine generalizing our results to more complex geometries, including manifolds with a non-uniform curvature. For an inhomogeneous but negatively curved everywhere space, we expect the main qualitative result about the absence of weak localization remains. A more interesting case is a mixed curvature space exhibiting a combination of positive (spherical) and negative (hyperbolic) curvature segments. This interestingly connects to a problem first considered by Zeldovich~\cite{zeldovich,Sokoloff.2015} in the context of general relativity, where he considered a ``Universe homogeneous in the mean'' but with a metric that fluctuates randomly
$$
g_{ab} = g^{(0)}_{ab} + \delta g_{ab},
$$
where $ g^{(0)}_{ab}$ is a flat Euclidean metric and $\delta g_{ab}$ corresponds to small fluctuations around flatness. Zeldovich has shown that the ``average'' Universe of such type will appear hyperbolic due to the phenomenon of intermittency (e.g., gravitational lensing will be observed as if the space is negatively curved)~\cite{Sokoloff.2015}. This suggests, per a similar argument, that mixed curvature manifolds with random curvature may avoid localization due to being effectively open, even if they are still flat on average. 
However a rigorous proof of this conjecture (tied to the problem of finding the return probability of a random walker on a mixed curvature manifold) appears to be a difficult mathematical problem. 


Finally, it may be possible to test our results experimentally. 
Recent breakthroughs in the engineering of quantum simulators have enabled the experimental design and construction of superconducting resonators which can be arranged in to a hyperbolic lattice structure~\cite{Kollar.2019,Lenggenhager.2022,Boettcher.2020,Boettcher.2022}.
Provided these can be manufactured with a tunable level of disorder and made large enough, these could actually realize the lattice version of the physics we consider here and confirm the presence or absence of disorder-induced localization. 
It is also possible that sufficiently thin two-dimensional materials like graphene may exhibit long-wavelength strain or surface roughness that effectively generates ``puddles" of regions with locally positive or negative spatial curvature~\cite{Wei.2023,Vozmediano.2008,Wood.2023}, connecting once again to the problem posed by Zeldovich.

\begin{acknowledgements}
The authors would like to acknowledge Alireza Parhizkar, Jeet Shah, Gautam Nambiar, and Alexey Gorshkov for productive discussions.
J.B.C. and P.N. acknowledge support from the Quantum Science Center (QSC), a National Quantum Information Science Research Center of the U.S. Department of Energy (DOE). V.G. was supported by the U.S. Department of Energy, Office of Science, Basic Energy Sciences under Award No. DE-SC0001911
\end{acknowledgements}

\bibliography{references}

\end{document}